\documentclass[aps,prb,twocolumn,showpacs,superscriptaddress,floatfix]{revtex4}  
\usepackage[english]{babel}
\usepackage{graphicx}
\usepackage{dcolumn}
\usepackage{amsmath}
\usepackage{natbib}
\usepackage{bm}

\begin{document}

\title{Circular dichroism of magneto-phonon resonance in doped graphene}

\author{P.~Kossacki}\email{Piotr.Kossacki@fuw.edu.pl}
\affiliation{Laboratoire National des Champs Magne\'{e}tiques
Intenses, CNRS/UJF/UPS/INSA, F-38042 Grenoble, France}
\affiliation{Institute of Experimental Physics, University of
Warsaw, Hoza 69, 00-681 Warsaw, Poland}

\author{C.~Faugeras}
\affiliation{Laboratoire National des Champs Magne\'{e}tiques
Intenses, CNRS/UJF/UPS/INSA, F-38042 Grenoble, France}

\author{M.~K\"{u}hne}
\affiliation{Laboratoire National des Champs Magne\'{e}tiques
Intenses, CNRS/UJF/UPS/INSA, F-38042 Grenoble, France}

\author{M.~Orlita}
\affiliation{Laboratoire National des Champs Magne\'{e}tiques
Intenses, CNRS/UJF/UPS/INSA, F-38042 Grenoble, France}

\author{A.~Mahmood}
\affiliation{CEMES-CNRS, Universit\`{e} de Toulouse, 29 rue Jeanne
Marvig, 31055 Toulouse, France } \affiliation{CNRS-Institut
N\'{e}el, BP 166, 38042 Grenoble Cedex 9, France }

\author{E.~Dujardin}
\affiliation{CEMES-CNRS, Universit\'{e} de Toulouse, 29 rue Jeanne
Marvig, 31055 Toulouse, France  }

\author{R.R.~Nair}
\affiliation{School of Physics and Astronomy and Centre for
Mesoscience and Nanotechnology, University of Manchester,
Manchester M13 9PL, UK}

\author{A.K.~Geim}
\affiliation{School of Physics and Astronomy and Centre for
Mesoscience and Nanotechnology, University of Manchester,
Manchester M13 9PL, UK}

\author{M.~Potemski}
\affiliation{Laboratoire National des Champs Magne\'{e}tiques
Intenses, CNRS/UJF/UPS/INSA, F-38042 Grenoble, France}

\date{\today}

\begin{abstract}
Polarization resolved, Raman scattering response due to E$_{2g}$
phonons in monolayer graphene has been investigated in magnetic
fields up to 29~T. The hybridization of the E$_{2g}$ phonon is
only observed with the fundamental inter Landau level excitation
(involving the n=0 Landau level) and in just one of the two
configurations of the circularly crossed polarized excitation and
scattered light. This polarization anisotropy of the
magneto-phonon resonance is shown to be inherent to relatively
strongly doped graphene samples, with carrier concentrations
typical for graphene deposited on Si/SiO$_2$ substrates.
\end{abstract}

\pacs{73.22.Lp, 63.20.Kd, 78.30.Na, 78.67.-n}
\maketitle

The effective coupling of optical phonons (E$_{2g}$) to electronic
excitations in graphene yields a particularly remarkable,
resonant, magneto-phonon effect if the Raman scattering response
of the E$_{2g}$ phonon (so called G-band) is investigated as a
function of the magnetic field $B$ applied across the layer. In
this case, the two-dimensional (2D) energy bands of graphene
condense into discrete Landau levels L$_{\pm n}$ with energies
$E_{\pm n} = \pm v_F \sqrt{2e\hbar Bn}$ ($n=0,  1, 2, ... $).
Then, the E$_{2g}$ phonon hybridizes with the selected
L$_{-n-1~(-n)}$~$\rightarrow$~L$_{n~(n+1)}$ inter Landau level
(LL) excitations~\cite{Ando07,Goerbig07}. As a consequence, a
series of avoided crossing events (in clean systems) and/or
noticeable magneto-oscillations in the phonon response (in more
disordered systems) can be seen in Raman scattering
experiments~\cite{Faugeras09,Yan2010,Faugeras2011}. So far, only
quasi-neutral graphene structures; epitaxial
graphene~\cite{Faugeras09} and "electronically decoupled" graphene
on a graphite surface~\cite{Yan2010,Faugeras2011,Kuehne2012}, have
shown a clear magneto-phonon resonant effect. The amplitude of
this effect depends on the electron-phonon coupling constant
$\lambda$ and the oscillator strength of the inter LL transition
involved, which includes matrix elements and occupation factors of
the initial and final state Landau levels. For neutral graphene,
where the Fermi level is pinned to the L$_0$ level, all inter-LL
excitations are active. Indeed this system shows a rich series of
avoided crossings events each time the
L$_{-n-1~(-n)}$~$\rightarrow$~L$_{n~(n+1)}$ transition approaches
the E$_{2g}$ mode. It is interesting to examine the case of doped
graphene where the Fermi energy is far away from the neutrality
point (L$_0$ level) and a part of the
L$_{-n-1~(-n)}$~$\rightarrow$~L$_{n~(n+1)}$ transitions become
inactive due to Pauli blocking. The latter implies the suppression
of their hybridization with the optical phonon. Polarization
resolved Raman scattering experiments are expected to be crucial
for tracking the magneto-phonon resonance in doped
graphene~\cite{Goerbig07}. Indeed, the two, degenerate in energy
L$_{-n}$~$\rightarrow$~L$_{n+1}$ and
L$_{-n-1}$~$\rightarrow$~L$_{n}$ transitions have identical
oscillator strength in neutral graphene but not necessarily so if
the Fermi energy is pushed away from the neutrality point.
L$_{-n}$~$\rightarrow$~L$_{n+1}$  and
L$_{-n-1}$~$\rightarrow$~L$_{n}$ excitations carry either m$_z$=+1
or -1 angular momentum. Their possible difference in hybridization
strength with the E$_{2g}$ phonon should be detectable with the
circular polarization of the incident and scattered
photons~\cite{Goerbig07}.

Here, we report on the progress in searching for a magneto-phonon
resonance in monolayer graphene on Si/SiO$_{2}$ substrate. We
employed the high-field magneto-Raman scattering methods and
applied polarization resolved measurements to investigate the
graphene's G-band in two relevant, $\sigma ^{+}/\sigma ^{-}$ and
$\sigma ^{-}/\sigma ^{+}$ configurations of circularly polarized
excitation/scattered light. The E$_{2g}$ phonon shows a pronounced
hybridization only with the fundamental inter LL transition
(involving the L$_{0}$-level) at a resonant field B$\approx25$~T
and, distinctly, only in the case of one of the two polarization
configurations. This behavior is shown to be characteristic for
graphene with carrier concentration of $\sim 2\times
10^{12}\textrm{cm}^{-2}$, typical for untreated samples of
graphene on Si/SiO$_{2}$. Such a relatively strong, either n- or
p-type, doping implies that all interband transitions involving
LLs with indexes $n>1$ are blocked due to the occupation factor.
Nevertheless at B~$\sim25$~T the LL filling factor is $2<v<6$, and
either L$_{0}$~$\rightarrow$~L$_{1}$ (electron doping) or
L$_{-1}$~$\rightarrow$~L$_{0}$ (hole doping), but not both of
them, become active. This explains our observation of the
pronounced "circular dichroism" in the magneto-phonon resonance of
doped graphene.

The magneto-Raman scattering experiments have been carried out in
back-scattering Faraday geometry using a homemade set-up -
see~Ref.\onlinecite{Kuehne2012} for details of technical
arrangements. The 488nm line of an Ar$^{+}$ ion laser has been
used as the excitation source\cite{Wang2008}. The laser light
(with power $\sim 1$~mW) has been focused on the sample, down to a
spot of $\sim 1\mu\textrm{m}$ in diameter. The sample has been
mounted on an XYZ piezo-stage. The measurements, in two distinct,
$\sigma ^{+}/\sigma ^{-}$ and $\sigma ^{-}/\sigma ^{+}$
configurations for the circular polarisation of the
incident/collected light were realized by inverting the direction
of the magnetic field, while keeping fixed the orientation of all
optical elements (e.g., polarizers). Experiments have been carried
out at liquid helium temperatures (T=4.2K) and in magnetic fields
up to 29T (supplied by a resistive magnet).

\begin{figure}
\includegraphics[width=0.43\textwidth]{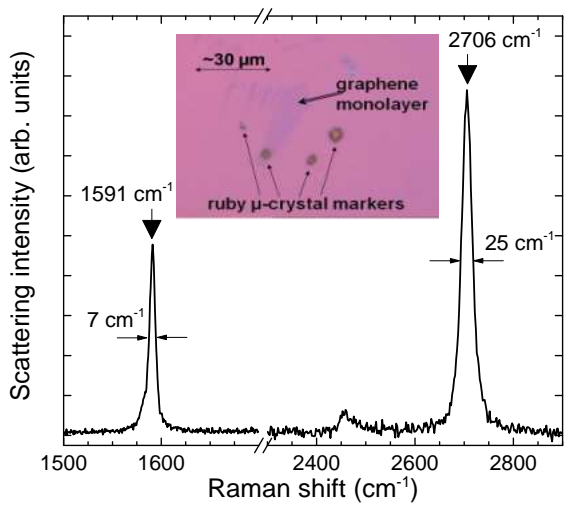}
\caption{\label{fig1} Raman scattering spectra measured at
temperature T=4.2~K and under 488~nm laser line excitation,
showing the characteristic E$_{2g}$ and 2D bands of the graphene
investigated. The inset shows the optical image of the sample. }
\end{figure}

The studied graphene sample was exfoliated from natural graphite
and deposited on top of a Si/SiO$_{2}$ wafer (oxide thickness of
$300$~nm). Its optical microscope image is shown in the inset of
Fig.~\ref{fig1}. Micron-size ruby crystals (giving rise to a
strong luminescence signal at $\sim$~692nm) have been deposited in
the vicinity of the graphene flake in order to facilitate the
localization of the sample with respect to the laser spot. Raman
scattering spectra measured in the absence of the magnetic field
and at temperature T=4.2~K confirm the monolayer character of the
graphene structure investigated \cite{Ferrari06,Graf07} (see
Fig.~\ref{fig1}). The so-called 2D band, centered at
$2706$~cm$^{-1}$, fairly represents a single Lorentzian line with
the characteristic Full Width at Half Maximum,
FWHM~$=25$~cm$^{-1}$ . The G-band shows the characteristic width
of FWHM~$=7$~cm$^{-1}$ and is centered at $1591$~cm$^{-1}$. Both
its small width and energy position point towards a considerable
(electron or hole) doping of the investigated graphene sample
\cite{Yan07,Pisana07}.

\begin{figure}
\includegraphics[width=0.35\textwidth]{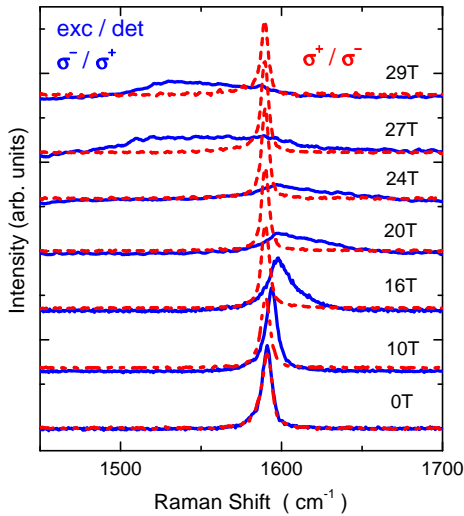}
\caption{\label{fig3} (Color online) Typical Raman scattering
spectra of the G band feature in the two crossed circular
polarization configurations (dashed line corresponds to
$\sigma^+/\sigma^-$) at various values of the magnetic field.
Spectra are shifted for clarity.}
\end{figure}

Our experimental findings are presented in Fig.~\ref{fig3}. This
figure illustrates the evolution with the magnetic field of the
measured magneto-Raman spectra of the E$_{2g}$ phonon in two,
distinct, $\sigma ^{-}/\sigma ^{+}$ and $\sigma ^{+}/\sigma ^{-}$
configurations of circularly polarized excitation/scattered light.
The prominent feature seen in this figure is a pronounced circular
magneto-dichroism of the Raman scattering spectra of the
investigated G-band. This band is practically unaffected by the
magnetic field when probed in one $\sigma ^{+}/\sigma ^{-}$
polarization configuration. Focusing more on the results obtained
in this configuration we find that the FWHM of the G-band does not
indeed change with the magnetic field up to the highest values
explored, of 29~T. The central position of the G-band shows,
however, a small downward (red) shift in energy of up to
$\sim$~2cm$^{-1}$ at B=29~T, the shift remaining smaller than the
linewidth. The evolution of the G-band with the magnetic field is
clearly much richer in the opposite $\sigma ^{-}/\sigma ^{+}$
configuration. In this case, a pronounced blue shift of the
spectral weight of the G-band, associated with noticeable spectral
broadening, is seen in the spectra in magnetic fields up to ~24~T.
The shift reaches ~20~cm$^{-1}$ and the spectrum extends over
~50~cm$^{-1}$. At B~$\sim$~25~T, the G-band practically disappears
from the spectra, likely being smeared over a very broad energy
range. At still higher fields, the G-band reappears in the spectra
as a broad line but this time on the low energy side of its
original position (at B=0). This characteristic evolution of the
$\sigma ^{-}/\sigma ^{+}$ component of the G-band points towards
its effective hybridization (avoided crossing) with one of the
L$_{-1}$~$\rightarrow$~L$_{0}$ or L$_{0}$~$\rightarrow$~L$_{1}$
transitions. At the same time, none of those two excitations
couple with the $\sigma ^{+}/\sigma ^{-}$ component of the G-band.
We therefore suspect that we are dealing with strongly doped
graphene, such that at high fields around B=25~T, the Landau level
filling factor is $2< \mid \nu \mid<6$. Then, in case of electron
doping ($\nu > 0$) the L$_{0}$ Landau level is fully populated
with electrons but L$_{1}$ is partially empty and
L$_{0}$~$\rightarrow$~L$_{1}$ transition is active but
L$_{-1}$~$\rightarrow$~L$_{0}$ is Pauli blocked. In case of hole
doping($\nu < 0$) the $L_{0}$ level is completely empty but
electrons occupy the $L_{-1}$ level and then the
L$_{-1}$~$\rightarrow$~L$_{0}$ transition is active but not
L$_{0}$~$\rightarrow$~L$_{1}$. In consequence, the E$_{2g}$ phonon
can couple with only one of possible (fundamental) transitions
involving the $L_{0}$ level, which broadly explains the strong
circular dichroism observed in our magneto-Raman scattering data.

To ascertain the light helicities of our polarization
configurations, we have replaced the graphene sample by a
semiconductor, GaAs/(Ga,Al)As quantum well structure and
calibrated the experimental set-up by measuring the near band-edge
magneto-luminescence of this structure with a well known sequence
of two, Zeeman-split, polarization-resolved
components~\cite{Potemski98}. Taking into account the
characteristic for graphene selection
rules~\cite{Goerbig07,Kossacki2011} resulting from the transfer of
angular momentum we deduce that in the $\sigma ^{-}/\sigma ^{+}$
configuration the E$_{2g}$ phonon couples to the
L$_{-1}$~$\rightarrow$~L$_{0}$ transition and therefore conclude
that our graphene is hole doped. An initial inspection of the data
points at the resonant field, when $E_{2g}=E_{1}$, of $ B \simeq
25T$. This is somewhat lower as compared to previous observations,
at B~$ \simeq 28$~T in case of quasi neutral epitaxial graphene
\cite{Sadowski06,Orlita08} and graphene on graphite
\cite{Yan2010,Faugeras2011}, and indicates a slightly higher Fermi
velocity ( $ \simeq 1.1\times10^{6} m/s$ instead of $ \simeq
1.0\times10^{6} m/s$) in the present case of graphene flake on
Si/SiO~$_{2}$ \cite{Jiang07,Deacon07}. We use this value of Fermi
velocity in our simulations.  We speculate that different
"effective" Fermi velocities in different samples might be also
accounted for by different degree of disorder but a discussion of
those effects is beyond the scope of this paper and will be
presented elsewhere.


With the aim to better support our data interpretation, we
illustrate in Fig.3 the three representative regimes (with respect
to carrier concentration) of the magneto-phonon resonance in
graphene. Hole doping is considered, however, the results hold for
n-type graphene as well but with the inverted polarization
configurations. Following the concepts developed in Refs
\onlinecite{Ando07,Goerbig07} we refer to the phonon Green
function :

\begin{equation}
\label{Equ1} \tilde{\epsilon}^{2}-\epsilon_{0}^{2} = 2
\epsilon_{0} \lambda E_{1}^{2} \sum _{k=0} ^{\infty} \{\frac{f_{k}
T_{k}}{(\tilde{\epsilon}+i\delta)^{2}-T_{k}^{2}}+\frac{1}{T_{k}}\}
\end{equation}

which defines the actual phonon energy $\epsilon  =
Re(\tilde{\epsilon})$ (and broadening $Im(\tilde{\epsilon})$), and
in which $\epsilon_{0}$ is the phonon energy of the neutral system
at $B=0$. $\delta$ accounts for the broadening characteristic for
electronic transitions. $T_k=(\sqrt{k}+\sqrt{k+1})E_{1}$,
($k=0,1,2, ... $), stands for energies of inter-Landau level
excitations which possibly couple to the phonon. $f_k$, which
varies from 0 to 1, is defined as the filling factor of the single
Landau level and accounts for the occupancy of the initial and
final states involved in a given
L$_{-n-1~(-n)}$~$\rightarrow$~L$_{n~(n+1)}$ excitation.

In the following, we consider only the phonon energy $\epsilon$,
and assume the same broadening of electronic states of $\delta =
90 \textrm{cm}^{-1}$ as in Ref. \onlinecite{Faugeras09} which
partially smears the effect of mode hybridization (jumps in energy
rather than clear anticrossings).  The phonon energy in two
different crossed polarization configurations is calculated by two
separate solutions of Eq.1 : the first taking into account only
the $\Delta |n|=+1$ inter Landau transitions
(L$_{-n}$~$\rightarrow$~L$_{n+1}$) and the second one accounting
separately for $\Delta |n|=-1$ excitations
(L$_{-n-1}$~$\rightarrow$~L$_{n}$).

\begin{figure}
\includegraphics[width=85mm]{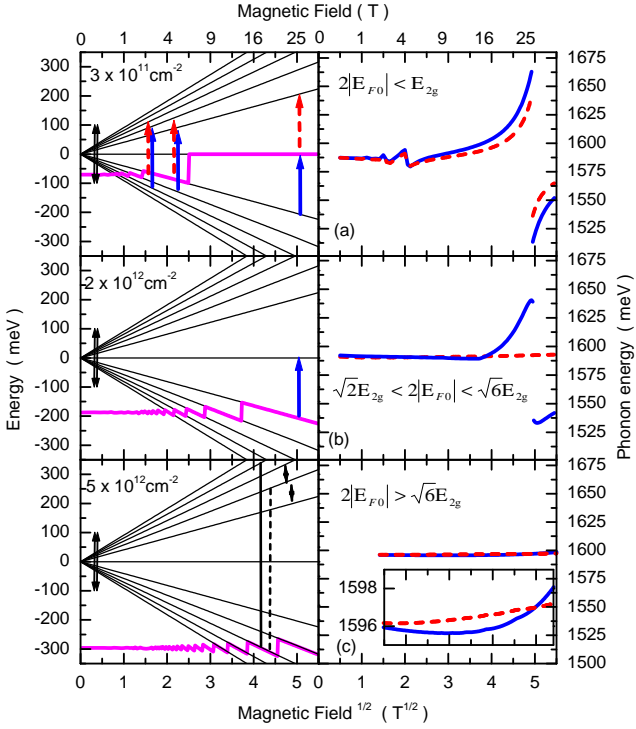}
\caption{\label{fig4} (Color online) Three regimes of the hole gas
concentration and its impact on the magneto-phonon effect. Left
panels represent fan charts of the Landau levels with marked
excitations being in resonance with $E_{2g}$ phonon (panels a and
b). The solid arrows represent excitations in $\sigma ^{-}/\sigma
^{+}$ polarization, dashed in $\sigma ^{+}/\sigma ^{-}$
polarization. The thick line shows the Fermi level. Arrows on
panel c represent non resonant excitations (inter- and
intra-band). Right panels show the evolution of the phonon energy
versus magnetic field, with the same representation of
polarizations by solid/ dashed lines. }
\end{figure}

Varying the carrier concentration, one may distinguish three
distinct regimes for the magneto-phonon resonance in graphene. The
first one appears at relatively low doping, when
$E_{2g}>2|E_{F0}|$ ($E_{F0}$ stands for the Fermi energy at zero
magnetic field). Then, practically all
L$_{-n-1~(-n)}$~$\rightarrow$~L$_{n~(n+1)}$ transitions are active
and effectively hybridize with the phonon (see two upper panels of
Fig.~\ref{fig4}). In this case the nonvanishing carrier
concentration results in a rather small asymmetry in the two
distinct polarization configurations, the details of the asymmetry
being determined by the actual evolution of the Fermi energy with
the magnetic field. The second regime is illustrated in two middle
panels of Fig.~\ref{fig4}. It reflects the situation when
$\sqrt{2} |E_{F0}|$ exceeds E$_{2g}$ and that all
L$_{-n-1~(-n+1)}$~$\rightarrow$~L$_{n~(n)}$ transitions with
$n\geq1$ are Pauli blocked at magnetic fields of their resonances
with the phonon. At the same time, the carrier concentration can
be low enough that at the field value when $E_{1}=E_{2g}$, the
Fermi energy is pinned to the L$_{-1}$ level ($2
|E_{F0}|<\sqrt{6}E_{2g}$) and therefore the
L$_{-1}$~$\rightarrow$~L$_{0}$ transition is active and it
resonantly hybridizes with the phonon. Notably this is the only
resonant hybridization event possible in this case which likely
corresponds to the present experimental results.

Apart of the pronounced anticrossing events, the $E_{2g}$ phonon
is also affected by interactions with other, more distant in
energy, inter LL excitations. This leads to an additional, though
much weaker and rather smooth evolution of the G-band with the
magnetic field. These non-resonant effects persist even in case of
high carrier concentrations where they appear to be particularly
transparent and can be estimated analytically. We discuss them
here in an attempt to account for the small red shift of the
phonon energy with the magnetic field, which is observed in the
inactive $\sigma ^{+}/\sigma ^{-}$ polarisation configuration. As
illustrated in the bottom panels of Fig.~\ref{fig4}, when the hole
concentration is high enough, i.e., when ($2
|E_{F0}|>\sqrt{6}E_{2g}$), all resonant hybridizations of the
E$_{2g}$ phonon with L$_{-n-1~(-n)}$~$\rightarrow$~L$_{n~(n+1)}$
excitations are quenched. Nevertheless, the G-band still displays
certain field dependence (and circular dichroism). To elucidate
this behavior, we firstly note that the non resonant
renormalization of the phonon energy which results from
interactions with all high energy $\Delta |n|=+1$ inter Landau
level transitions is not the same as the renormalization resulting
from the interaction with all $\Delta |n|=-1$ excitations. To see
this, imagine that the Fermi level is located low in the band, in
between $L_{-l}$ and $L_{-l+1}$ level. Then, there is clearly one
more active transition (L$_{-l}$~$\rightarrow$~L$_{l-1}$)  in the
series of L$_{-n}$~$\rightarrow$~L$_{n-1}$ transitions as compared
to the series of L$_{-n}$~$\rightarrow$~L$_{n+1}$
(L$_{-l+1}$~$\rightarrow$~L$_{l}$ is blocked). Due to this effect,
the estimated splitting of the phonon energy (in two different
polarization configurations) is $\epsilon_{-} - \epsilon_{+} =
\lambda E_{CR}$ where  $E_{CR}=v_F e B / \sqrt{\pi n_{h}} =
E_1^2/(2|E_{F0}|)$. Secondly, we have so far neglected a possible
coupling of the phonon with the (low energy) cyclotron resonance
like mode which occurs between neighboring Landau levels in the
vicinity of the Fermi level. Notably, this mode, present only in
one polarisation, may also couple to the phonon \cite{Ando07},
leading to the estimated blue shift of the phonon energy of about
$\lambda \frac{2 v_F^{3} \hbar e^2 B^2}{ {\epsilon_0}^2 \sqrt{\pi
n_{h}} }$, in the range of magnetic fields considered.
Interestingly, an additional resonant hybridization event which
involves this cyclotron resonance
mode~\cite{Witowski2010,Orlita2012} may be characteristic of
strongly doped graphene but arising at fields much above the range
explored in these experiments, i.e., at $B=E_{2g}\sqrt{\pi
n_{h}}/v_F e$ if $\nu = \sqrt{n_{h}/\pi}hv_F/E_{2g} \gg 1$. For
example for carrier density $n = 10^{13}cm^{-2}$ the resonance
would occur at B$\sim$110T. The results of numerical calculations
of phonon energies in two polarization configurations for graphene
with high hole concentration ($n_{h}=5\times 10^{12}cm^{-1}$) are
shown in the lowest panel of Fig.~\ref{fig4}. The non resonant
effects are overall weak but nonetheless the expected trends are
against the observation of the noticeable red shift of the phonon
experienced in the "inactive" $\sigma ^{+}/\sigma ^{-}$
configuration. This observation remain a puzzle and requires more
elaborated theoretical models, perhaps including the effects of
possible electron hole asymmetry of the graphene bands and or
invoking different coupling mechanism with the cyclotron resonance
mode.

\begin{figure}
\includegraphics[width=0.35\textwidth]{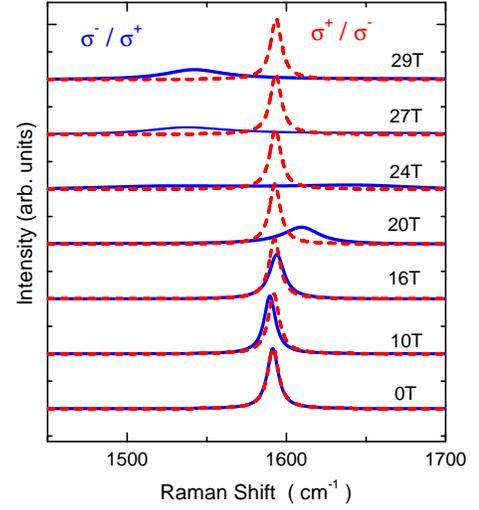}
\caption{\label{fig5} Simulation of the Raman spectra versus
magnetic field in two crossed circular polarization configurations
for different values of the magnetic field.}
\end{figure}

 The above analysis
was done in the ideal case of a well defined, uniform carrier
density and for a graphene system with no other scattering than
the electronic scattering introduced through the parameter
$\delta$ in Eq.~\ref{Equ1}. Despite their simplicity our
simulations help us to clarify the main expected features of the
magneto-phonon resonances in graphene in different possible
regimes of carrier concentration.
 The spectra presented in
Fig.~\ref{fig3} show no evolution of the phonon feature in one
polarization configuration and a clear anti-crossing behavior at
$B=25T$ in the other polarization. This observation indicates that
$|E_{F0|}
> \sqrt{2} E_{2g} / 2$ ($n_{h}=1.2\times
10^{12}~\textrm{cm}^{-2}$) and $|E_{F0}| < \sqrt{6} E_{2g} / 2$
($n_{h}=3.5\times 10^{12}~\textrm{cm}^{-2}$). Hence, we are
investigating the second density regime, for which a single
inter-band electronic excitation, i.e., the
L$_{-1}$~$\rightarrow$~L$_{0}$ transition is active when tuned in
resonance to the phonon energy which results in the observed
anticrossing effect.

The qualitative attempt to simulate the shape of the measured
spectra is presented in Fig.~\ref{fig5}. The simulations imply the
calculation of the phonon spectral function, $-\textrm{Im}
D(q,\omega)$, following the procedure described by
Ando~\cite{Ando07}(Eqs. 2.21 and 3.2) complemented with the
associated polarization selection rules~\cite{Goerbig07}. We have
set the parameter $\lambda$ to $4.5 \times 10^{-3}$ in line with
its value derived in Ref.~[\onlinecite{Faugeras09}]. Each spectrum
is convoluted with a Gaussian function with a FWHM of
$6.6~\textrm{cm}^{-1}$, in order to reproduce the experimentally
observed FWHM of the G-band at zero magnetic field. The parameter
$\delta = 90~\textrm{cm}^{-1}$ and the carrier density
$n_{el}=2\times 10^{12}~\textrm{cm}^{-2}$ were adjusted to fairly
reproduce the experimental shift of the phonon feature in the
range of magnetic fields prior to the anticrossing event observed
in $\sigma ^{-}/\sigma ^{+}$ polarization configuration. The
assumed value of the hole concentration implies a complete
quenching of the anticrossing event in the opposite polarization
configuration. As mentioned before, the amplitude of the phonon
energy shift strongly depends on $\lambda$  - a hypothesis of a
smaller value for $\lambda$ would imply an assumption of a smaller
value for $\delta$.

With the assumed values of relevant parameters, the main
experimental features are fairly well simulated: the phonon
feature is almost unchanged up to $29$T in the $\sigma ^{+}/\sigma
^{-}$ polarization, while in the $\sigma ^{-}/\sigma ^{+}$
configuration the pronounced anticrossing effect, at B$\simeq25$~T
is reproduced. The main differences between simulation and the
experiment are: (i) the smooth development of the blue shift
without any distinguishable kink, and (ii) the asymmetric
line-shape observed in the experiment close to the resonant
magnetic field. These discrepancies, we believe are due to our
simplified approach in accounting for the effects of
disorder~\cite{Kashuba2012}. A more detailed analysis would
require taking into account the fluctuations of local carrier
density and possibly an exact form of Landau level broadening.
Nevertheless we conclude that the experimental observations are
overall well reproduced by our calculated spectra.

To conclude, by using two different crossed circular polarization
configurations, we have been able to investigate the coupling of
the E$_{2g}$ phonon either to $\Delta |n|=+1$ or $\Delta |n|=-1$
inter Landau level excitations. The strong anisotropy in the
magneto-phonon effect observed in two distinct  polarization
configurations at high magnetic field is accounted for by
relatively high doping of the investigated graphene. The
hybridization of the E$_{2g}$ phonon is effective with the
L$_{-1}$~$\rightarrow$~L$_{0}$ transition whereas Pauli blocking
of the L$_{0}$~$\rightarrow$~L$_{1}$ transition quenches the
phonon coupling with this latter electronic excitation. More
investigations, in particular on gated graphene structures are
needed to uncover all the subtleties of the magneto-phonon
resonance in graphene.

\begin{acknowledgments}
Part of this work has been supported by GACR P204/10/1020,
GRA/10/E006 (EPIGRAT), RTRA "DISPOGRAPH" projects and by
EuroMagNET II under the EU contract number 228043. One of us (PK)
acknowledges the support from European Project No.
FP7/2007-2013-221515 (MOCNA).
\end{acknowledgments}



\end{document}